\input{aipcheck}

\documentclass[
    ,final            
  ]
  {aipproc}

\usepackage{amsmath}
\usepackage{amssymb}
\usepackage{graphics}
\usepackage{graphicx}
\usepackage{psfrag}
\usepackage[mathcal]{eucal}
\layoutstyle{6x9}

\begin{document}

\title{Models of Neutrino Masses and Mixing\footnote{Based on talk presented at the conference on Colliders to Cosmic Rays 2007 (C2CR07), Lake Tahoe, CA, February 25 - March 1, 2007.}}


\author{Mu-Chun Chen}{
  address={Department of Physics \& Astronomy, University of California, Irvine, CA 92697-4575\\
  and\\
  Theoretical Physics Department, Fermilab, Batavia, IL 60510-0500},
  email={mcchen@fnal.gov}
}

\begin{abstract}
Neutrino physics has entered an era of precision measurements. With these precise measurements, 
we may be able to distinguish different models that have been constructed to explain the small 
neutrino masses and the large mixing among them.  In this talk, I review some of the existing theoretical models and their predictions for neutrino oscillations. 
\end{abstract}

\maketitle


\section{Introduction}

The recent  advent of the neutrino oscillation data from Super-Kamiokande has provided a solid evidence that neutrinos have small but non-zero masses. The global fit to current data from neutrino oscillation experiments give the following best fit values and $2\sigma$ limits for the mixing parameters~\cite{Maltoni:2004ei},
\begin{eqnarray}
\sin^{2} \theta_{12} & = & 0.30 \; (0.25 - 0.34), \quad \Delta m_{12}^{2} = 7.9 \; (7.3 - 8.5) \; \mbox{eV}\\
\sin^{2} \theta_{23} & = &  0.5 \; (0.38 - 0.64), \quad \Delta m_{23}^{2} = 2.2 \; (1.7 - 2.9) \; \mbox{eV}\\
\sin^{2} \theta_{13} & = & 0 \;  (< 0.028) \; .
\end{eqnarray}
Since then, the measurements of neutrino oscillation parameters have entered a precision era. In the Standard Model, due to the lack of  right-handed neutrinos and lepton number conservation, neutrinos are massless. To generate non-zero neutrino masses thus calls for physics beyond the Standard Model. There have been many theoretical ideas proposed with an attempt to accommodate the experimentally observed  small neutrino masses and the larger mixing angles among them. Most of the models are either based on grand unification combined with family symmetries or having family symmetries in the lepton sector only.  Recently, it was realized that small neutrino masses can also arise with new physics at the TeV scale, contrary to the common belief that the scale of the seesaw mechanism has to be high. 
In this talk, I review some of these ideas as well as the predictions of various existing models. For more extensive reviews, see, for example, Ref.~\cite{Chen:2003zv}.

\section{Models with Discrete Family Symmetries}

These values for the mixing parameters are very close to the values arising from the so-called ``tri-bimaximal'' mixing (TBM) matrix~\cite{Harrison:2002er},
\begin{equation}
U_{\mathrm{TBM}} = \left(\begin{array}{ccc}
\sqrt{2/3} & 1/\sqrt{3} & 0\\
-\sqrt{1/6} & 1/\sqrt{3} & -1/\sqrt{2}\\
-\sqrt{1/6} & 1/\sqrt{3} & 1/\sqrt{2}
\end{array}\right) \; , \label{eq:tri-bi}
\end{equation}
which predicts $\sin^{2}\theta_{\mathrm{atm, \, TBM}} = 1/2$ and $\sin\theta_{13, \mathrm{TBM}} = 0$. In addition, it predicts $\sin^{2}\theta_{\odot, \mathrm{TBM}} = 1/3$ for the solar mixing angle. Even though the predicted $\theta_{\odot, \mathrm{TBM}}$ is currently still  allowed by the experimental data at $2\sigma$, as it is very close to the upper bound at the $2\sigma$ limit, it  may be ruled out once more precise measurements are made in the  upcoming experiments.  It has been pointed out that the tri-bimaximal mixing matrix can arise from a family symmetry in the lepton sector based on $A_{4}$~\cite{Ma:2001dn}. However, due to its lack of doublet representations, CKM matrix is an identity in most $A_{4}$ models.   It is hence not easy to implement $A_{4}$ as a family symmetry for both quarks and 
leptons~\cite{Ma:2006sk}.

In \cite{Chen:2007af}, a grand unified model based on SU(5) combined with the double tetrahedral group~\cite{Frampton:1994rk}, ${}^{(d)}T$ was presented, 
which successfully, for the first time, gives rise to near tri-bimaximal leptonic mixing as well as realistic CKM matrix elements for the quarks. The charge assignments of various fields are summarized in Table~\ref{tbl:charge}.  
 \begin{table}
\begin{tabular}{|c|ccc|ccc|cccccc|cc|}\hline
& $T_{3}$ & $T_{a}$ & $\overline{F}$ & $H_{5}$ & $H_{\overline{5}}^{\prime}$ & $\Delta_{45}$ & $\phi$ & $\phi^{\prime}$ & $\psi$ & $\psi^{\prime}$ & $\zeta$ & $N$ & $\xi$ & $\eta$  \\ [0.3em] \hline\hline
SU(5) & 10 & 10 & $\overline{5}$ & 5 &  $\overline{5}$ & 45 & 1 & 1 & 1 & 1& 1 & 1 & 1 & 1\\ \hline
${ }^{(d)}T$ & 1 & $2$ & 3 & 1 & 1 & $1^{\prime}$ & 3 & 3 & $2^{\prime}$ & $2$ & $1^{\prime\prime}$ & $1^{\prime}$ & 3 & 1 \\ [0.2em] \hline
$Z_{12}$ & $\omega^{5}$ & $\omega^{2}$ & $\omega^{5}$ & $\omega^{2}$ & $\omega^{2}$ & $\omega^{5}$ & $\omega^{3}$ & $\omega^{2}$ & $\omega^{6}$ & $\omega^{9}$ & $\omega^{9}$ 
& $\omega^{3}$ & $\omega^{10}$ & $\omega^{10}$ \\ [0.2em] \hline
$Z_{12}^{\prime}$ & $\omega$ & $\omega^{4}$ & $\omega^{8}$ & $\omega^{10}$ & $\omega^{10}$ & $\omega^{3}$ & $\omega^{3}$ & $\omega^{6}$ & $\omega^{7}$ & $\omega^{8}$ & $\omega^{2}$ & $\omega^{11}$ & 1 & $1$ 
\\ \hline   
\end{tabular}
\vspace{0.1in}
\caption{Charge assignments. Here the parameter $\omega = e^{i\pi/6}$.}  
\label{tbl:charge}
\end{table}
Due to the presence of the  
$Z_{12} \times Z_{12}^{\prime}$ symmetry, only nine operators are allowed in the model, and hence the model is very predictive, the total number of parameters being nine in the Yukawa sector for the charged fermions and the neutrinos. The Lagrangian of the model is given as follows,
\begin{eqnarray}
\mathcal{L}_{\mathrm{Yuk}} & = &  \mathcal{L}_{\mathrm{TT}} + \mathcal{L}_{\mathrm{TF}} + \mathcal{L}_{\mathrm{FF}} \\
\mathcal{L}_{\mathrm{TT}} & = & y_{t} H_{5} T_{3}T_{3} + \frac{1}{\Lambda^{2}} y_{ts} H_{5} T_{3} T_{a} \psi \zeta + 
\frac{1}{\Lambda^{2}} y_{c} H_{5} T_{a} T_{a} \phi^{2} + \frac{1}{\Lambda^{3}} y_{u} H_{5} T_{a} T_{a} \phi^{\prime 3}   \\  
\mathcal{L}_{\mathrm{TF}} & = & \frac{1}{ \Lambda^{2}}  y_{b} H_{\overline{5}}^{\prime} \overline{F} T_{3} \phi \zeta + \frac{1}{\Lambda^{3}} \biggl[ y_{s} \Delta_{45} \overline{F} T_{a} \phi \psi N  + 
 y_{d} H_{\overline{5}}^{\prime} \overline{F} T_{a} \phi^{2} \psi^{\prime}  \biggr]\\
\mathcal{L}_{\mathrm{FF}} & = & \frac{1}{M_{x}\Lambda} \biggl[\lambda_{1} H_{5} H_{5} \overline{F} \, \overline{F} \xi +  \lambda_{2} H_{5} H_{5} \overline{F} \, \overline{F} \eta\biggr] \; ,
 \end{eqnarray}
 where $M_{x}$ is the cutoff scale at which the lepton number violation operator $HH\overline{F}\, \overline{F}$ is generated, while $\Lambda$ is the cutoff scale, above which the ${}^{(d)}T$ symmetry is exact.  The parameters $y$'s and $\lambda$'s are the coupling constants. 
Due to the $Z_{12}$ symmetry,  the mass hierarchy arises dynamically without invoking additional U(1) symmetry. Due to the ${ }^{(d)}T$ transformation property of the matter fields, the $b$-quark mass can be generated only when the ${ }^{(d)}T$ symmetry is broken, which naturally explains  the hierarchy between $m_{b}$ and $m_{t}$. 
The $Z_{12} \times Z_{12}^{\prime}$ symmetry, to a very high order, also forbids operators that lead to nucleon decays. In principle, a symmetry smaller than  $Z_{12} \times Z_{12}^{\prime}$ would suffice in getting realistic masses and mixing pattern; however, more operators will be allowed and the model would not be as predictive.  
The Georgi-Jarlskog relations for three generations are obtained. This inevitably requires non-vanishing mixing in the charged lepton sector, leading to correction to the tri-bimaximal mixing pattern. The model predicts non-vanishing $\theta_{13}$, which is related to the Cabibbo angle as, $\theta_{13}\sim \theta_{c}/3\sqrt{2}$.  In addition, it gives rise to a sum rule, $\tan^{2}\theta_{\odot} \simeq \tan^{2} \theta_{\odot, \mathrm{TBM}} - \frac{1}{2} \theta_{c} \cos\beta$,  which is a consequence of the Georgi-Jarlskog relations in the quark sector. This deviation could account for the difference between the experimental best fit value for the solar mixing angle and the value predicted by the tri-bimaximal mixing matrix.

\subsection{Models with GUT Symmetries}

The non-zero neutrino masses give support to the
idea of grand unification based on $SO(10)$ in which all the 16 fermions
(including the right-handed neutrinos) can be accommodated in one single spinor
representation. Furthermore, it provides a framework in which the seesaw
mechanism arises naturally. Models based on $SO(10)$ combined with a 
continuous or discrete flavor symmetry group have been constructed 
to understand the flavor problem, especially the small neutrino masses and 
the large leptonic mixing angles. These models can be classified 
according to the family symmetry implemented as well as the Higgs representations introduced in the model. 
Phenomenologically, the resulting mass matrices can be either symmetric, lop-sided, or asymmetric.

Due to the product rule, $16 \otimes 16 = 10 \oplus 120_{a}  \oplus 126_{s}$, 
the only Higgses that can couple to the matter fields at tree level are in the 10, $\overline{120}$, and $\overline{126}$ representations of SO(10). The Yukawa matrices involving the 10 and $\overline{126}$ are symmetric under interchanging the family indices while that involving the $\overline{120}$ is anti-symmetric. The Majorana mass term for the RH neutrinos can arise either from a renormalizable operator involing $\overline{126}$, or from a non-renormalizable operator that involves the $16$'s. The case of $\overline{126}$ has the advantage that R-parity is preserved automatically. 

If $SO(10)$ is broken through the left-right 
symmetric breaking route, the resulting fermion mass matrices are symmetric. In this case, both the large solar mixing 
angle and the maximal atmospheric mixing angle come from  
the effective neutrino mass matrix. A characteristic of this class of models is that 
the predicted value for $|U_{e\nu_{3}}|$ element tends to be larger than the value 
predicted by models with lopsided textures~\cite{Albright:2001uh}. This GUT symmetry breaking pattern gives rise to the following relations among various mass matrices,
\begin{equation}
M_{u} = M_{\nu_{D}}, \quad M_{d} = M_{e} \; ,
\end{equation}
up to some calculable group theoretical factors which are useful in obtaining the Jarlskog relations among masses for the charged leptons and down type quarks, when combined with family symmetries. 
 The value of $U_{e3}$ is predicted to be large, 
close to the sensitivity of current experiments. The prediction for the rate of $\mu \rightarrow e\gamma$ is about two orders 
of magnitude below the current experimental bound.

In a particular model constructed in~\cite{Chen:2000fp}, the Higgs sector contains fields in 10, 45, 54, 126 representations,
with 10, 126 breaking the EW symmetry and generating fermions masses, and 
45, 54, 126 breaking the SO(10) GUT symmetry. 
The mass hierarchy can arise if there is an $SU(2)_{H}$ symmetry acting non-trivially 
on the first two generations such that the first two generations transform as a doublet and 
the third generation transforms as a singlet under $SU(2)_{H}$, which breaks down at two steps, 
$   SU(2) {\epsilon M \atop \rightarrow} U(1) 
{\epsilon^{\prime} M \atop \rightarrow} \mbox{nothing}$ where $   \epsilon^{\prime} 
\ll \epsilon \ll 1$. 
The mass hierarchy is generated 
by the Froggatt-Nielsen mechanism.
The resulting mass matrices at the GUT scale are given by
\begin{equation}  {
M_{u,\nu_{LR}}=
\left( \begin{array}{ccc}
  {0} & 
  {0} & 
  {\left<10_{2}^{+} \right> \epsilon'}\\
  {0} & 
  {\left<10_{4}^{+} \right> \epsilon} & 
  {\left<10_{3}^{+} \right> \epsilon} \\
  {\left<10_{2}^{+} \right> \epsilon'} & \
  {\left<10_{3}^{+} \right> \epsilon} &
  {\left<10_{1}^{+} \right>}
\end{array} \right)
= 
\left( \begin{array}{ccc}
  {0} & 
  {0} & 
  {r_{2} \epsilon'}\\
  {0} & 
  {r_{4} \epsilon} & 
  {\epsilon} \\
  {r_{2} \epsilon'} & 
  {\epsilon} & 
  {1}
\end{array} \right) M_{U}} \; ,
\end{equation}
\begin{equation}  {
M_{d,e}=
\left(\begin{array}{ccc}
  {0} & 
  {\left<10_{5}^{-} \right> \epsilon'} & 
  {0} \\
  {\left<10_{5}^{-} \right> \epsilon'} &  
  {(1,-3)\left<\overline{126}^{-} \right> \epsilon} & 
  {0}\\ 
  {0} & 
  {0} & 
  {\left<10_{1}^{-} \right>}
\end{array} \right)
=
\left(\begin{array}{ccc}
  {0} & 
  {\epsilon'} & 
  {0} \\
  {\epsilon'} &  
  {(1,-3) p \epsilon} & 
  {0}\\
  {0} & 
  {0} & 
  {1}
\end{array} \right) M_{D}} \; .
\end{equation}
The right-handed neutrino mass matrix is of the same form as $   M_{\nu_{LR}}$ 
\begin{equation}  {
M_{\nu_{RR}}=  
\left( \begin{array}{ccc}
  {0} & 
  {0} & 
  {\left<\overline{126}_{2}^{'0} \right> \delta_{1}}\\
  {0} & 
  {\left<\overline{126}_{2}^{'0} \right> \delta_{2}} & 
  {\left<\overline{126}_{2}^{'0} \right> \delta_{3}} \\ 
  {\left<\overline{126}_{2}^{'0} \right> \delta_{1}} & 
  {\left<\overline{126}_{2}^{'0} \right> \delta_{3}} &
  {\left<\overline{126}_{1}^{'0} \right> }
\end{array} \right)
= 
\left( \begin{array}{ccc}
  {0} & 
  {0} & 
  {\delta_{1}}\\
  {0} & 
  {\delta_{2}} & 
  {\delta_{3}} \\ 
  {\delta_{1}} & 
  {\delta_{3}} & 
  {1}
\end{array} \right) M_{R} \; .
\label{Mrr}}
\end{equation}
Note that, since we use
the $\overline{126}$-dimensional Higgs representation to generate the heavy
Majorana neutrino mass terms, R-parity is preserved at all energies. 
The effective neutrino mass matrix is 
\begin{equation}  {
\label{eq:Mll}
M_{\nu_{LL}}=M_{\nu_{LR}}^{T} M_{\nu_{RR}}^{-1} M_{\nu_{LR}}
= \left( 
\begin{array}{ccc}  
  {0} &   {0} &   {t} \\
  {0} &   {1} &   {1+t^{\prime}} \\  
  {t} &   {1+t^{\prime}} &   {1} 
\end{array} \right) \frac{d^{2}v_{u}^{2}}{M_{R}}} 
\end{equation}
giving rise to maximal mixing angle for the atmospheric neutrinos and LMA solution 
for the solar neutrinos. The form of the neutrino mass matrix in this model 
is invariant under the seesaws mechanism. 
The value of $U_{e3}$ is related to the ratio $\sim \sqrt{\Delta m_{sol}^{2}/\Delta m_{atm}^{2}}$, which is predicted to be  
close to the sensitivity of current experiments. 
The prediction for the rate of $\mu \rightarrow e\gamma$ is about two orders 
of magnitude below the current experimental bound~\cite{Chen:2004xy}. 

\section{TeV Scale Seesaw Mechanism}

In the conventional wisdom, the smallness of the neutrino masses is tied to the high scale of the new physics that generates neutrino masses. As the new physics scale is high, it is very hard, if not impossible, to probe such new physics at current collider experiments. In \cite{Chen:2006hn}, an alternative was proposed in which the small the neutrino masses are generated with TeV scale physics. This allows the possibility of testing the new physics that gives rise to neutrino masses at the Tevatron and the LHC. This is achieved by augmenting the Standard Model with a non-anomalous $U(1)_{\nu}$ symmetry and right-handed neutrinos. Due to the presence of  the $U(1)_{\nu}$ symmetry, neutrino masses can only be generated by operators with very high dimensionality, which in turn allows a low cut-off scale.

The new anomaly cancellation conditions
are highly non-trivial, especially because all fermion
charges are expected to be commensurate. Nevertheless, assuming that all quark Yukawa couplings
and all diagonal charged-lepton Yukawa couplings to the standard model Higgs doublet $H$ are gauge invariant,
 it is found that the most general solution to the anomaly cancellation conditions when
$N=1$ or 3. Only in the  $N=3$ case, scenarios consistent with
light neutrino masses and
$\Lambda$ at the TeV scale were found. 
For $N=3$, the charges of all quarks and leptons
(including right-handed neutrinos) are determined in terms of four rational parameters,
assumig one of the fermion charges is fixed by an appropriate normalization of the gauge
coupling. 

There exist regions in the Leptocratic Model  parameter space that fit the neutrino oscillation data.
Depending on the choice of parameters, the neutrinos can be  either
Dirac or Majorana fermions.
In scenarios with Majorana neutrinos, the existence of
``quasi-sterile'' neutrinos that mix slightly with the active neutrinos and couple to the new
$Z^{\prime}$ gauge boson is predicted. These quasi-sterile neutrinos may have interesting
phenomenological consequences 
for cosmology and oscillation physics. In the case of Dirac neutrinos, 
potentially observable consequences of the new degrees of freedom are also predicted.

Because the $U(1)_{\nu}$ symmetry is spontaneously broken around the weak scale, 
the $Z^{\prime}$ gauge boson and the particles from the $U(1)_{\nu}$ breaking sector
will manifest themselves in a variety of interesting ways.  
$Z^{\prime}$ exchange can mediate neutral-fermion flavor
violating processes, which may be observable in next-generation
neutrino oscillation experiments. The new heavy states can be discovered
in current and upcoming collider experiments, such as the
Tevatron, LHC and ILC, enabling the possibility of probing the neutrino sector at the collider experiments. 

\section{Predictions for the Oscillation Parameters}

\begin{figure}
 \includegraphics[height=.4\textheight]{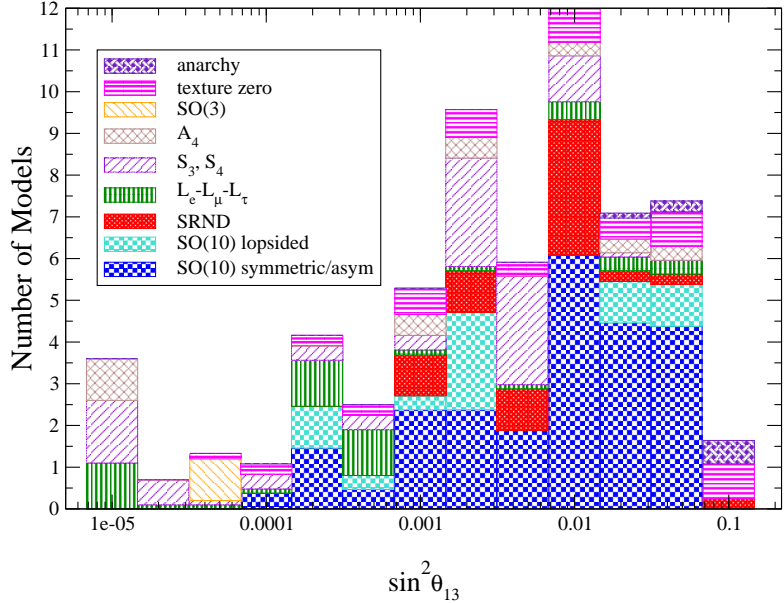}
 \caption{Predictions for $\theta_{13}$ from various models.}\label{fig:all}
\end{figure}

In \cite{Albright:2006cw}, a comparison of the predictions of some sixty-three models was presented, 
These include models based on $SO(10)$, 
models that utilize single RH neutrino dominance mechanism, 
and models based on family symmetries such as $L_{e} - L_{\mu} - L_{\tau}$ symmetry, 
$S_{3}$ symmetry, 
$A_{4}$ symmetry, 
and $SO(3)$ symmetry, 
as well as models based on texture zero assumptions. 
The predictions of these models for $\sin^{2}\theta_{13}$ are summarized in Fig.~\ref{fig:all}.    An observation one can draw immediately is that predictions of  $SO(10)$ models are larger than $10^{-4}$, and the median value is roughly $\sim 10^{-2}$.  Furthermore, $\sin^{2}\theta_{13} < 10^{-4}$ can only arise in models based on leptonic symmetries. However, these models are not as predictive as the GUT models, due to the uncertainty in the charged lepton mixing matrix. In this case, to measure $\theta_{13}$ will require a neutrino superbeam or a neutrino factory. 

\section{Cosmological Connection -- Leptogenesis}

The evidence of non-zero neutrino masses opens up the possibility that the leptonic CP violation might be responsible, through leptogenesis, for the observed asymmetry between matter and anti-matter in the Universe. (For a recent review, see, for example, Ref.~\cite{Chen:2007fv}.) It is generally difficult, however, to make connection between leptogenesis and CP-violating processes at low energies due to the presence of extra phases and mixing angles in the right-handed neutrino sector. Recently attempts have been made to induce {\it spontaneous CP violation}  (SCPV)  from a single source.  In a minimal left-right symmetric model proposed in \cite{Chen:2004ww}, SCPV could be due to two intrinsic CP violating phases associated with VEVs of two scalar fields, which account for all CP-violating processes observed in Nature; these {\it exhaust} sources of CP-violation. As the left-handed (LH) and right-handed (RH) Majorana mass matrices are identical up to an overall mass scale, in this model there exist relations between low energy processes, such as neutrino oscillations, neutrinoless double beta decay and lepton flavor violating charged lepton decay, and leptogenesis which occurs at very high energy~\cite{Chen:2004ww}. To yield a sufficient amount of baryonic asymmetry, the leptonic Jarlskog 
invariant $J_{CP}^{\ell}$ has to be greater than $10^{-5}$ in this model.  By imposing an additional $U(1)$ symmetry, the $SU(2)_{R} \times U(1)_{B-L}$ breaking scale, and thus the seesaw scale,  
can be made to be much lower compared to the GUT scale while still naturally giving rise to small neutrino masses with 
all coupling constants assuming natural values.  
In this case, there also exists strong relation between CP violation in the leptonic sector and CP violation in the quark 
sector~\cite{Chen:2006bv}. With the seesaw scale being $10^{3}$ TeV, the electric dipole moment of the electron is predicted in this  model to be $\sim 10^{-32}$ e-cm, which is accessible to the next generation of experiments.

\begin{figure}
 \includegraphics[height=.4\textheight,angle=270]{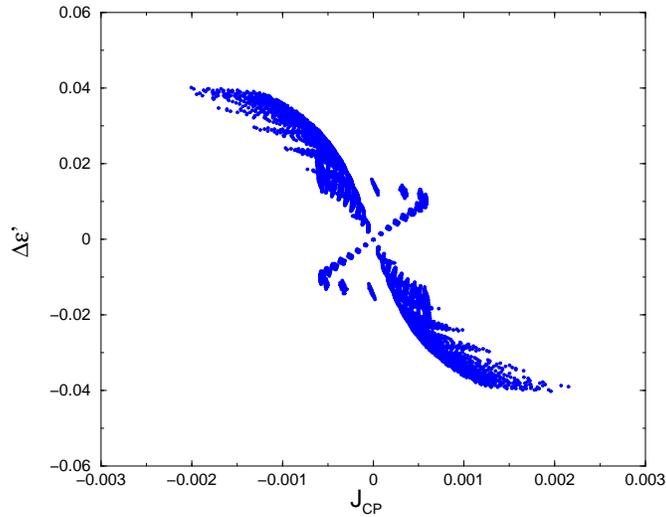}
 \caption{Correlation between lepton number asymmetry and the leptonic Jarlskog invariant.}
\end{figure}

\section{Conclusion}

In this talk, I have reviewed a few existing models for neutrino masses and mixing. In particular, I have presented a successful recent attempt based on a SU(5) grand unified model combined with ${}^{(d)}T$ symmetry, in which both the tri-bimaximal neutrino mixing and realistic CKM mixing matrix are generated.    A model in which small neutrino masses are generated with new physics at the TeV scale has also been shown.   A study of some sixty-three existing models indicates that the range of predictions of these models for $\theta_{13}$ is very broad, although there are some characteristic model predictions with which more precise experimental measurements may tell different models apart.  And finally, a very predictive framework based on minimal left-right symmetry with spontaneous CP violation is presented, where very strong correlations between leptogenesis and low energy CP violation processes can be established.  



\begin{theacknowledgments}
 It is a pleasure to thank the organizers, especially Bob Svoboda and Mani Tripathi, for the kind invitation and for the hospitality they extended during the workshop. The work of M-CC was supported, in part, by the start-up funds from the University of California at Irvine. 
 \end{theacknowledgments}



\bibliographystyle{aipproc}   




\end{document}